\PassOptionsToPackage{table}{xcolor}
\documentclass[cameraready]{Interspeech}

\usepackage{amsmath,graphicx}
\usepackage[fontsize=9pt]{scrextend}

\usepackage{booktabs}
\usepackage{balance}

\usepackage{amsmath,empheq, epsfig,amssymb,amsthm,url,amsfonts}
\usepackage{algorithm}
\usepackage{algpseudocode}
\usepackage{multirow}
\usepackage{mdframed}
\usepackage{url}
\usepackage{enumitem}
\usepackage[makeroom]{cancel}
\usepackage{dblfloatfix}
\usepackage{array}
\usepackage{comment}
\usepackage{epstopdf}
\usepackage{float}
\usepackage{subfig}
\usepackage{breqn}
\usepackage{cases}
\usepackage{caption}
\usepackage{tabularx}
\usepackage[printonlyused,withpage]{acronym}
\usepackage{balance}
\usepackage{graphbox} 

\usepackage{hyperref}
\usepackage{placeins}

\usepackage{xcolor}
\usepackage{colortbl}

\definecolor{gray}{rgb}{0.9,0.9,0.9} 
\usepackage{tikz}
\usetikzlibrary{fit,calc}

\colorlet{mypink}{red!40}
\colorlet{myblue}{cyan!60}

\setlist[itemize]{label=$\triangleright$}
\newtheoremstyle{break}
{}
{}
{\itshape}
{}
{\bfseries}
{.}
{\newline}
{}
\theoremstyle{break}

\theoremstyle{definition}

\usepackage{caption}



\newcommand{\vect}[1]{\mathbf{#1}}
\newcommand{\bs}[1]{\boldsymbol{#1}}
\newcommand{\E}{\mathbb{E}}
\usepackage{url}

\makeatletter
\def\thmhead@plain#1#2#3{%
	\thmname{#1}\thmnumber{\@ifnotempty{#1}{ }\@upn{#2}}%
	\thmnote{ {\the\thm@notefont#3}}}
\let\thmhead\thmhead@plain
\makeatother

\graphicspath{{./figs/}}

\newcommand{\argmax}{\operatornamewithlimits{argmax}}

\newcommand{\lk}{ \left\{ }
\newcommand{\rk}{ \right\} }

\newcommand{\diag}{\mbox{{diag}}}

\newsavebox\mybox

\acrodef{SE}{Speech enhancement}
\acrodef{CE}{Cross-Entropy}
\acrodef{AVSE}{audio-visual speech enhancement}
\acrodef{STFT}{short-time Fourier transform}
\acrodef{ESTOI}{extended short-time objective intelligibility}
\acrodef{STOI}{short-time objective intelligibility}
\acrodef{NMF}{non-negative matrix factorization}
\acrodef{DNN}{deep neural network}
\acrodef{VAE}{variational auto-encoder}
\acrodef{DKF}{deep Kalman filter}
\acrodefplural{VAEs}{variational auto-encoders}
\acrodef{EM}{expectation-maximization}
\acrodef{TF}{time-frequency}
\acrodef{ELBO}{evidence lower bound}
\acrodef{LR}{Living Room}
\acrodef{SDR}{signal-to-distortion ratio}
\acrodef{PESQ}{perceptual evaluation of speech quality}
\acrodef{SNR}{signal-to-noise ratio}
\acrodef{DNNs}{deep neural networks}
\acrodef{NN}{neural network}
\acrodef{VESDE}{variance-exploding stochastic differential equation}
\acrodef{SDE}{stochastic differential equation}
\acrodef{GAN}{generative adversarial networks}
\acrodefplural{GANs}{generative adversarial networks}
\acrodef{SI-SDR}{scale-invariant signal-to-distortion ratio}
\acrodef{SI-SAR}{scale-invariant signal-to-artefact ratio}
\acrodef{SI-SIR}{scale-invariant signal-to-interference ratio}
\acrodef{MOS}{mean opinion score}
\acrodef{SGMSE+}{score-based generative model for speech enhancement}
\acrodef{NCSNPP++}{Noise-Conditional Score Network}
\acrodef{WSJ}{Wall Street Journal}
\acrodef{UDiffSE}{Unsupervised Diffusion-Based Speech Enhancement}
\acrodef{PC}{Predictor-Corrector}
\acrodef{DDPM}{Denoising Diffusion Probabilistic Modeling}
\acrodef{DMPS}{Diffusion Model Posterior Sampling}

\newcommand{\romain}{\textcolor{green}}

\usepackage{bm}

\usepackage{cite}
\usepackage{textcomp}
\def\BibTeX{{\rm B\kern-.05em{\sc i\kern-.025em b}\kern-.08em
    T\kern-.1667em\lower.7ex\hbox{E}\kern-.125emX}}

\title{Audio-visual Contrastive Alignment for Diffusion-based Visual-conditioned Speech Enhancement\thanks{This work was supported by the French National Research Agency (ANR) under the project REAVISE (ANR-22-CE23-0026-01).}}

\author{Colombe}{Mboungou}
\author{Mostafa}{Sadeghi}
\author{Jean-Eudes}{Ayilo}
\author{Romain}{Serizel}

\address{
    Université de Lorraine, CNRS, Inria, Loria, Nancy, France
}

\email{\{colombe.mboungou, mostafa.sadeghi, jean-eudes.ayilo\}@inria.fr, romain.serizel@loria.fr}

\keywords{audio-visual speech enhancement, diffusion models, contrastive learning, robustness, low SNR.}

\usepackage{comment}

\begin{document}

\maketitle

\begin{abstract}

Audio-visual speech enhancement (AVSE) exploits visual cues such as lip movements to recover speech in noisy environments. Recent work introduced diffusion-based unsupervised AVSE, where a speech diffusion model conditioned on visual features via cross-attention is trained and used as a data-driven prior for posterior sampling-based speech enhancement. Despite promising performance over its audio-only counterpart, the impact of explicitly enforcing cross-modal alignment in the fusion remains unclear. In this work, we propose to augment the diffusion training objective with a contrastive audio–visual loss to encourage stronger use of visual information while keeping the posterior sampling framework unchanged. Experiments across matched and mismatched test data show consistent improvements in interference suppression, signal reconstruction, and perceptual quality, with the largest gains at low SNRs.
Code is available at \url{https://github.com/cexauce/AV-CA-DiffUSE}
\end{abstract}


\section{Introduction}

Speech enhancement (SE) aims to recover clean speech from noisy observations, 
a long-standing problem in speech processing that remains challenging 
under low signal-to-noise ratio (SNR) conditions and non-stationary noise. 
Recent deep learning approaches have significantly improved performance~\cite{pascual2017segan, luo2019conv, hu2020dccrn}, 
with generative modeling frameworks emerging as a powerful direction. 
In particular, score-based generative models such as SGMSE+~\cite{richter2023speech} have demonstrated 
that diffusion processes can effectively model the distribution of clean speech 
and achieve strong SE performance.

Audio-visual speech enhancement (AVSE) extends this paradigm by leveraging 
visual cues such as lip movements ~\cite{alfouras2018conversation, hou2018audio, jung2024flowavse}, which provide complementary and 
noise-robust information about speech content. 
Building on diffusion-based SE, recent unsupervised methods such as AV-UDiffSE+ ~\cite{ayilo2024diffavse}
leverage a pre-trained speech prior model conditioned on visual features and demonstrate robust generalization
performance across challenging noise conditions. In this approach, SE is performed by diffusion-based posterior sampling guided by a non-negative matrix factorization (NMF)-based observation model.

Despite these advances, the effectiveness of AVSE systems largely depends 
on how audio and visual modalities are fused ~\cite{michelsanti2021overview}. Most existing approaches, particularly unsupervised methods ~\cite{ayilo2024diffavse}, rely 
on architectural conditioning mechanisms without explicitly enforcing 
cross-modal representation alignment. However, multimodal contrastive 
learning frameworks such as CLIP~\cite{radford2021clip} and CLAP~\cite{elizalde2023clap} have shown that alignment 
objectives can produce rich and transferable representations.

Motivated by these findings, we investigate whether explicitly encouraging 
audio-visual alignment during diffusion-based pretraining can strengthen 
cross-modal fusion for unsupervised AVSE. Rather than introducing a separate pretraining 
stage\cite{qian2025sav}, we integrate a contrastive alignment objective directly into the score 
model optimization, allowing dynamic balancing between generative 
reconstruction and representation alignment, while keeping the inference-time algorithm intact. Our experiments confirm that imposing audio-visual embedding alignment enhances visual conditioning, leading to improved AVSE performance compared to the baseline \cite{ayilo2024diffavse},
particularly under low-SNR and cross-dataset mismatch scenarios. 

\section{Diffusion-based unsupervised AVSE}
We briefly review the diffusion-based unsupervised SE baseline, AV-DiffUSEEN, which combines the audio-visual training pipeline of AV-UDiffSE+~\cite{ayilo2024diffavse} with an extended unsupervised SE inference scheme, called DiffUSEEN~\cite{ayilo2026diffusion}. 

The SE problem is considered in the \ac{STFT} domain, where the observed noisy speech $\mathbf{x} \in \mathbb{C}^{FL}$ is modeled as
$\mathbf{x} = \mathbf{s} + \mathbf{n} + \mathbf{r}$. In this mixture, 
$\mathbf{s}$, $\mathbf{n} \in \mathbb{C}^{FL}$ are, respectively, the clean speech and the interfering noise STFTs, with $F$ frequency bins and $L$ time frames. The variable $\mathbf{r} \sim \mathcal{N}_{\mathbb{C}}(\bs{0}, \sigma_r^2\vect{I})$ is a Gaussian noise added to ease derivation with $\sigma_r$ being a small fixed constant. The clean speech $\mathbf{s}$ is associated to the visual features $\mathbf{v}$ extracted from the speaker’s lip movements. The interfering noise is supposed to be a Gaussian with a covariance that can be decomposed with non-negative matrix factorization (NMF) i.e. $\vect{n} \sim \mathcal{N}_{\mathbb{C}}(\bs{0}, \diag(\vect{m}_{\phi}))$, where $\vect{m}_{\phi}=\text{vec}(\vect{W}\vect{H})$, with $\text{vec}(.)$ the vectorization operator, $\vect{W}$ and $\vect{H}$ being low-rank matrices with non-negative entries and form the noise parameters $\phi=\{\vect{W},\vect{H}\}$. To estimate the clean speech, unsupervised SE methods~\cite{ayilo2024diffavse,ayilo2026diffusion} employ a Bayesian inference framework, in which a score-based diffusion model is trained as a clean-speech prior and then used at inference time to perform posterior sampling of the clean speech.

\subsection{Conditional score-based speech prior}
The score-based diffusion model~\cite{song2021scorebased} used in \cite{ayilo2024diffavse} learns to estimate the gradient of the conditional log-density of clean speech data, i.e., the score function $\nabla_{\mathbf{s}} \log p(\mathbf{s}|\mathbf{v})$. To this end, a forward stochastic differential equation (SDE) is used to perturbate the clean speech with Gaussian noise at different scale over a time horizon $t\in (0,1]$ and a neural network is trained to approximate the score function $\nabla_{\mathbf{s}_t} \log p_t(\mathbf{s}_t|\mathbf{v})$ of the diffused clean speech $\mathbf{s}_t$. Concretely, the forward SDE writes:
\begin{equation}\label{eqn:sde-fwd}
    \textrm{d}\vect{s}_t = \vect{f}(\vect{s}_t) \textrm{d}t + g(t) \textrm{d}\vect{w},
\end{equation}
where $\vect{f}(\vect{s}_t) = -\gamma \vect{s}_t$, $\gamma>0$ being a constant parameter. Here, $\vect{w}$ represents a standard Wiener process, $\vect{u}$ denotes the drift coefficient, and $g(t)$ is the diffusion coefficient, controlling the noise scale. Furthermore, the forward SDE admits a reverse SDE allowing to sample clean speech from the prior distribution $p(\mathbf{s}|\mathbf{v})$ by progressively transforming a Gaussian noise into a clean speech conditionally to the visual feature $\mathbf{v}$. It is formulated as follows:
\begin{equation}\label{eqn:rev-sde}
    \textrm{d}\vect{s}_t = \big[\vect{f}(\vect{s}_t) - g(t)^2 \nabla_{\vect{s}_t} \log p_t(\vect{s}_t|\vect{v})\big] \textrm{d}t + g(t) \textrm{d}\overline{\vect{w}},
\end{equation}
where $\overline{\vect{w}}$ is a standard Wiener process running backward in time, and $\textrm{d}t$ is a negative time increment.
The intractability of $\nabla_{\mathbf{s}_t} \log p_t(\mathbf{s}_t|\mathbf{v})$ makes the reverse SDE not directly usable and for that, a neural network is used to approximate it. Thus, a score network $\vect{S}_{\theta}$ parameterized by $\theta$ is trained with the weighted denoising score matching loss: 
\begin{equation}\label{eqn:ldiff}
    \mathcal{L}_{\textrm{Gen}}= \mathbb{E}_{t, \vect{s_0},\vect{v}, \bs{\zeta}} 
    \Big[\| \sigma(t)\vect{S}_{\theta}(\vect{s}_t, t, \vect{v}) + \bs{\zeta} \|_2^2\Big],
\end{equation}
where $\bs{\zeta}$ is the Gaussian noise added to the clean speech with a variance determined by the scaling parameter $\sigma(t)$. Once the score network is learned, $\vect{S}_{\theta}(\vect{s}_t, t, \vect{v})$ approximates $\nabla_{\vect{s}_t} \log p_t(\vect{s}_t|\vect{v})$ in equation \eqref{eqn:rev-sde} and the PC sampler \cite{song2021scorebased} is used to run the reverse SDE and sample the clean speech. 
\subsection{Unsupervised enhancement with DiffUSEEN}
The inference algorithm of {DiffUSEEN}~\cite{ayilo2026diffusion} applies an iterative Expectation-Maximisation (EM)-based algorithm to estimate the clean speech and the interfering noise parameters $\phi$. Concisely, the E-step samples the clean speech via diffusion-based posterior sampling, while the interfering noise is drawn from its conditional posterior. The M-step updates the noise parameters $\phi$ with the multiplicative update rules \cite{fevotte2009nonnegative}. The clean speech is sampled from the posterior $p(\vect{s}|\vect{v}, \vect{x}, {\vect{n}}) \propto p(\vect{x}|\vect{s}, \vect{n}) p(\vect{s}|\vect{v})$ by running the reverse SDE: 
\begin{equation}\label{eqn:rev-avdiffuseen}
\begin{split}
\textrm{d}\vect{s}_t = \Big[\vect{f}(\vect{s}_t) - g(t)^2 \nabla_{\vect{s}_t} \log p(\vect{x}|\vect{s}_t, \vect{n}) \\ - g(t)^2 \nabla_{\vect{s}_t} \log p_t(\vect{s}_t|\vect{v})\Big] \textrm{d}t + g(t) \textrm{d}\overline{\vect{w}},
\end{split}    
\end{equation}
where the posterior score function $\nabla_{\vect{s}_t} \log p_t(\vect{s}_t|\vect{v},\vect{x}, \vect{n}) = \nabla_{\vect{s}_t} \log p(\vect{x}|\vect{s}_t, \vect{n}) + \nabla_{\vect{s}_t} \log p_t(\vect{s}_t|\vect{v})$ replaces the prior score function $\nabla_{\vect{s}_t} \log p_t(\vect{s}_t|\vect{v})$ in \eqref{eqn:rev-sde}. 
In this work, we retain DiffUSEEN as the inference algorithm and focus on improving the audio-visual prior score network through contrastive alignment.

\section{Audio-visual contrastive alignment}

\begin{figure}[t]
  \centering
  \includegraphics[width=1.07\linewidth]{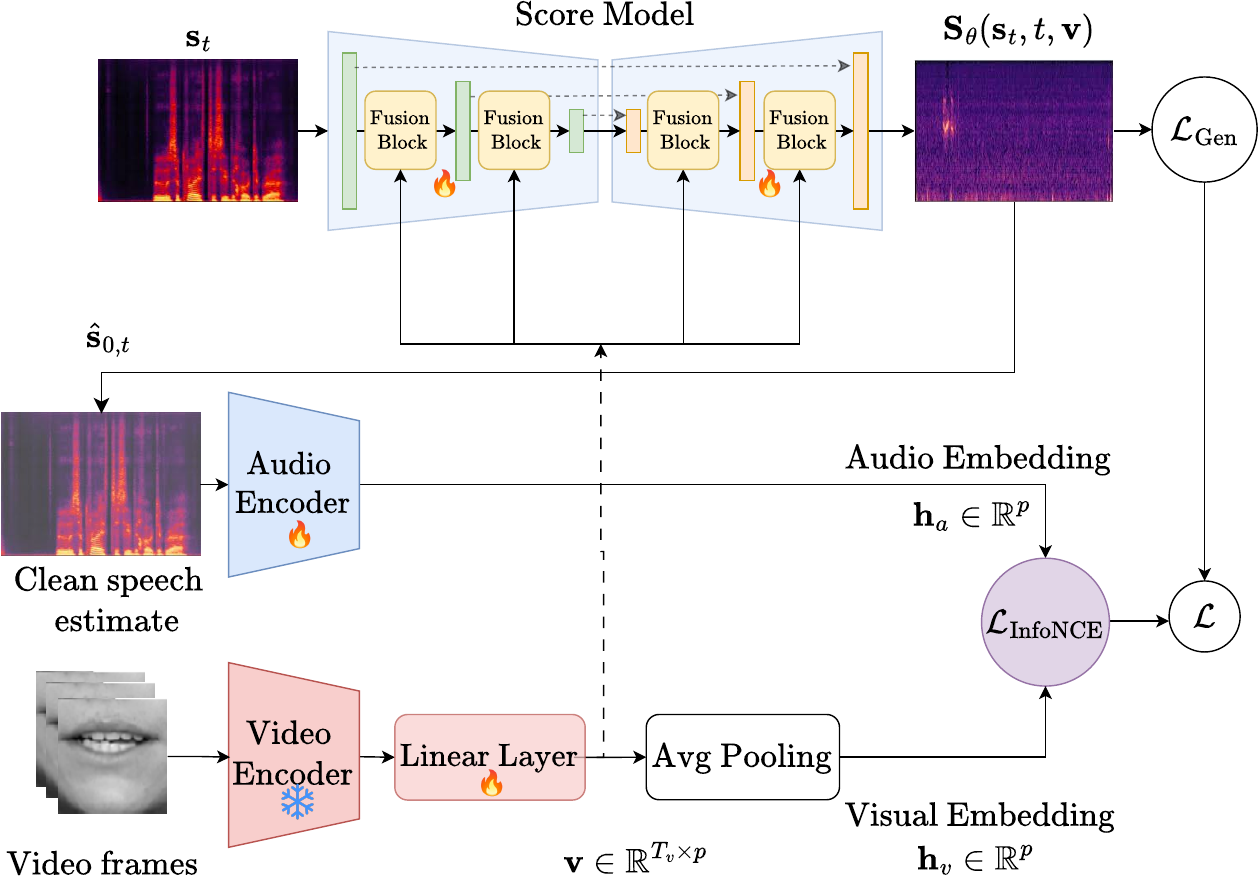}
  \caption{Audio-visual score model with contrastive alignment.}
  \label{fig:AV-UDiffSE with audio-visual contrastive alignment}
\end{figure}

This section presents our proposed contrastive alignment framework, depicted in Fig.~\ref{fig:AV-UDiffSE with audio-visual contrastive alignment}, describing how the audio-visual contrastive loss is computed and the motivation behind it.
\subsection{Limitations of baseline audio-visual fusion}
In AV-UDiffSE+~\cite{ayilo2024diffavse}, visual features are injected into the score network via a cross-attention mechanism, which conditions the audio representations on the visual stream. While this design allows the model to selectively exploit visual cues to improve speech reconstruction, it primarily optimizes for local reconstruction error at each diffusion step. As a result, cross-attention does not explicitly enforce global structure or alignment in any joint audio-visual embedding space. Consequently, the model may under-utilize visual information when the audio signal alone provides a strong prior. This motivates the introduction of an explicit contrastive audio-visual alignment objective, described in the following subsection, which encourages richer and more globally consistent audio-visual representations.

\subsection{Contrastive-augmented training objective}

\noindent\textbf{Clean speech estimation.}
Given the diffused (noisy) speech sample $\vect{s}_t$ at diffusion time-step $t$, we derive an estimate of the
clean speech using Tweedie's formula~\cite{efron2011tweedie}:
\begin{equation}\label{eq:tweedie}
\hat{\vect{s}}_{0,t} = \E_{p(\vect{s}|\vect{s}_t)}[\vect{s}] \approx
\frac{\vect{s}_t + \sigma^2_t \vect{S}_\theta(\vect{s}_t,\vect{v}, t)}{\textrm{e}^{-\gamma t}},
\end{equation}
where $\vect{S}_\theta$ denotes the U-Net-based score estimator.

\vspace{0.3cm}
\noindent\textbf{Audio and visual embeddings.}
The estimated clean speech $\hat{\vect{s}}_{0,t}$ is encoded by a
trainable ResNet-18~\cite{he2015resnet} audio encoder, denoted by $E_a$,
yielding $\vect{h}_a = E_a(\hat{\vect{s}}_{0,t})\in\mathbb{R}^p$. The visual input sequence $\vect{V} \in \mathbb{R}^{W \times H \times T_v}$
is encoded using a frozen pretrained AV-HuBERT~\cite{shi2022learning}
encoder, denoted by $E_v$. A learnable linear projection layer $W_v$ and a projection bias $b_v$ are applied to
match the conditioning space $\vect{v} = W_v E_v(\vect{V}) + b_v$. A temporal average pooling is applied on $\vect{v}$ to obtain the final visual embedding $\vect{h}_v$ used for the contrastive loss.

\vspace{0.3cm}
\noindent\textbf{Contrastive alignment objective.}  
We define the similarity between the $i$-th audio embedding and the $j$-th visual embedding (after applying unit-norm normalization) as \(s_{ij} = (\vect{h}_a^{(i)})^\top \vect{h}_v^{(j)} / \tau\), and use the symmetric InfoNCE~\cite{oord2019infonce} loss:
\begin{equation}
\mathcal{L}_{\textrm{InfoNCE}}
= -\frac{1}{2B} \sum_{i=1}^{B} 
\Bigg[
\log \frac{\exp(s_{ii})}{\sum_j \exp(s_{ij})} +
\log \frac{\exp(s_{ii})}{\sum_j \exp(s_{ji})}
\Bigg],
\end{equation}
where 
$\tau$ is a temperature parameter and $B$ is the batch size.

\vspace{0.3cm}
\noindent\textbf{Final training objective.}
The final loss combines the diffusion loss
$\mathcal{L}_{\textrm{Gen}}$ with the contrastive loss for each batch:
\begin{equation}
\mathcal{L}
=
\mathcal{L}_{\textrm{Gen}}
+
\alpha(t)\,
\beta(\textrm{epoch})\,
\mathcal{L}_{\textrm{InfoNCE}}.
\end{equation}
The schedule $\beta(\textrm{epoch})$ increases from 0 to $\beta_0$ over training epochs. Since the generative and contrastive losses differ in magnitude, the weighting hyperparameter $\beta_0$ is introduced to balance the two objectives. Conceptually, it controls the trade-off between speech reconstruction and audio-visual instance discrimination. The time-dependent weight $\alpha(t)$ is set to $1$ for $t \leq 0.3$ and $0$ for $t > 0.3$. This applies the alignment term only in the early denoising steps, where the Tweedie estimate in~\eqref{eq:tweedie} is sufficiently reliable, while letting the generative objective dominate as the process converges.





\subsection{A mutual information perspective}\label{sec:mi}

The proposed audio-visual contrastive loss can be viewed as encouraging stronger statistical dependence between the learned audio and visual representations. By increasing similarity for matched audio-visual pairs and decreasing it for mismatched pairs within a batch, the InfoNCE objective promotes cross-modal consistency and is often interpreted as maximizing a tractable lower-bound surrogate of mutual information between the two modalities. In practice, this encourages the denoised clean-speech estimates in \eqref{eq:tweedie} to remain compatible with the visual input, reinforcing the contribution of complementary visual cues during diffusion-based reconstruction. This perspective also clarifies potential failure modes: if the contrastive term is overweighted, the model may prioritize cross-modal agreement over acoustic fidelity, weakening the score-learning objective and degrading speech quality. Conversely, if the visual input is replaced by random vectors, the contrastive signal becomes uninformative and may either be ignored by the model or induce degenerate alignment behavior that reduces meaningful audio variation. Overall, these considerations motivate carefully balancing the diffusion loss with the contrastive alignment term.

\section{Experimental setup}
We analyze how contrastive cross-modal alignment affects interference suppression and robustness in diffusion-based AVSE across input SNR and data mismatch. As baselines, we consider AV-DiffUSEEN, with cross-attention fusion~\cite{ayilo2024diffavse,ayilo2026diffusion}, the audio-only version, AO-DiffUSEEN~\cite{ayilo2026diffusion}, and the supervised-generative FlowAVSE model \cite{jung2024flowavse}.

We conduct controlled ablation experiments in which (i) we study the influence of the contrastive weight and (ii) the visual embedding is replaced by a randomly sampled vector. These experiments allow us to empirically validate the failure modes discussed in Subsection~\ref{sec:mi} and highlight the importance of properly balancing cross-modal alignment with generative reconstruction. Results are presented in Section~5.
\subsection{Datasets}
Training was conducted on the TCD-TIMIT~\cite{harte2015tcdtimit} corpus, which contains clean audio-visual speech recorded in a controlled studio environment with frontal face recordings and high-quality audio. For evaluation, we consider TCD-DEMAND, obtained by mixing speech from TCD-TIMIT with environmental noise from the DEMAND~\cite{thiemann2013diverse} corpus. This represents a matched condition with unseen noise types. Moreover, to evaluate cross-dataset generalization, we consider LRS3-NTCD~\cite{ayilo2024diffavse}, created by mixing LRS3 speech data extracted from TED and TEDx videos recorded in unconstrained conditions~\cite{Afouras2018LRS3TEDAL} with NTCD noise~\cite{abdelaziz2017ntcd}. This results in a fully mismatched condition with unseen speakers and unseen visual and acoustic environments. In both cases, the SNR levels are $\{-5, 5\}$~dB. FlowAVSE was trained on TCD-DEMAND at SNR levels $\{-10, 0, 10\}$~dB, so the noise dataset is matched for this baseline.
FlowAVSE. It's probably worth mentionning}

\begin{table*}[!t]
\setlength{\tabcolsep}{4pt}
\renewcommand{\arraystretch}{0.95}
\centering
\caption{Average SE metrics in matched (TCD speech + DEMAND noise) and mismatched (LRS3 speech + NTCD noise) conditions. AO and AV: audio-only and audio-visual models, respectively. \textbf{Best value}, \underline{next best} per column.} 
\resizebox{0.98\linewidth}{!}{%
\begin{tabular}{lcccccccccc}
\toprule
                                 & \multicolumn{5}{c}{TCD speech + DEMAND noise} & \multicolumn{5}{c}{LRS3 speech + NTCD noise} \\
                                  \cmidrule(lr){2-6} \cmidrule(lr){7-11} 
\multicolumn{1}{l}{Method} 
& SI-SDR $\uparrow$ 
& SI-SIR $\uparrow$ 
& SI-SAR $\uparrow$ 
& PESQ $\uparrow$ 
& STOI $\uparrow$ 
& SI-SDR $\uparrow$ 
& SI-SIR $\uparrow$ 
& SI-SAR $\uparrow$ 
& PESQ $\uparrow$ 
& STOI $\uparrow$ \\
\midrule
Input 
& 0.00 & 0.00 & 55.7 & 2.83 & 0.70 
& 0.03 & 0.00 & 44.30 & 2.10 & 0.58 \\ 
\midrule
AO-DiffUSEEN ~\cite{ayilo2026diffusion} 
& 10.70 & 17.00 & 15.0 & 3.17 & 0.76 
& 5.83 & 8.95 & \textbf{10.00} & 2.44 & 0.65 \\ 
AV-DiffUSEEN ~\cite{ayilo2024diffavse}
& 13.60 & 24.30 & 15.6 & \textbf{3.28} & \underline{0.79} 
& \underline{7.40} & 15.0 & \underline{9.68} & \underline{2.58} & \textbf{0.68} \\
FlowAVSE~\cite{jung2024flowavse} (Supervised) 
& \textbf{17.80} & \textbf{39.9} & \textbf{17.90} & 3.18 & \textbf{0.82} 
& 3.12 & \textbf{21.2} & 3.28 & 1.49 & 0.53 \\
\midrule
Our model 
& \underline{16.0} & \underline{29.5} & \underline{16.10} & \textbf{3.28} & \underline{0.79} 
& \textbf{8.10} & \underline{18.60} & 9.50& \textbf{2.60} & \textbf{0.68} \\
\bottomrule
\end{tabular}
}
\label{tab:new_se_results_clean}
\end{table*}
\subsection{Evaluation metrics}
We evaluate the SE performance with some standard objective metrics: signal-to-distortion ratio (SI-SDR)~\cite{le2019sdr}, signal-to-interference ratio (SI-SIR)~\cite{vincent2006performance}, and signal-to-artifacts ratio (SI-SAR)~\cite{vincent2006performance}, all in dB, perceptual evaluation of speech quality (PESQ)~\cite{rix2001pesq} $[-0.5, 4.5]$, and short-time objective intelligibility (STOI)~\cite{taal2011stoi} $[0, 1]$.

SI-SDR measures overall reconstruction quality by quantifying the ratio between the target signal and the residual distortion after optimal scaling. SI-SIR evaluates the suppression of interference sources, whereas SI-SAR reflects the ratio between the level of the target speech signal and the level of artifacts introduced by the enhancement process. Higher values indicate better performance for all three metrics.

PESQ estimates perceived speech quality by modeling human auditory perception, whereas STOI correlates with speech intelligibility by measuring short-time temporal envelope similarity. Higher PESQ and STOI values indicate improved perceptual quality and intelligibility, respectively.
\subsection{Model architecture}
The baseline, AV-DiffUSEEN \cite{ayilo2024diffavse}, has 6.8~M parameters and uses an NCSN++M backbone, a lighter U-Net-like variant of NCSN++~\cite{richter2023speech}, with audio-visual fusion via single-head cross-attention modules. Our model extends AV-DiffUSEEN with a trainable ResNet-18 audio encoder (used only at training time), jointly optimized with the U-Net model and trained from scratch using an InfoNCE loss for audio-visual alignment, plus a linear projection layer on the AV-HuBERT embeddings before fusion. FlowAVSE employs a larger U-Net backbone with 60.2~M parameters.
\begin{table}[t]
\centering
\caption{TCD-DEMAND performance by input SNR.}
\label{tab:tcd_demand_full_snr}
\footnotesize
\setlength{\tabcolsep}{1.5pt}  
\begin{tabular}{l c c c c c c c c c c c}
\toprule
Model 
& \multicolumn{2}{c}{SI-SDR$\uparrow$} 
& \multicolumn{2}{c}{SI-SIR$\uparrow$} 
& \multicolumn{2}{c}{SI-SAR$\uparrow$} 
& \multicolumn{2}{c}{PESQ$\uparrow$} 
& \multicolumn{2}{c}{STOI$\uparrow$} \\
\cmidrule(lr){2-3} \cmidrule(lr){4-5} \cmidrule(lr){6-7} \cmidrule(lr){8-9} \cmidrule(lr){10-11}
& -5 & 5 & -5 & 5 & -5 & 5 & -5 & 5 & -5 & 5 \\
\midrule
AV-DiffUSEEN \cite{ayilo2024diffavse} & 10.1 & 17.0 & 21.2 & 27.4 & 12.12 & 19.1& 2.94 & 3.50 & 0.70 & 0.87 \\

Our Model & \textbf{13.3} & \textbf{18.7} & \textbf{27.8}& \textbf{31.2} & \textbf{13.7} & \textbf{19.3} & \textbf{3.0} & \textbf{3.56} & \textbf{0.72} & 0.87 \\

\bottomrule
\end{tabular}
\end{table}
\subsection{Hyper-parameter Settings}
For the baselines, we use the default hyper-parameters suggested in the associated papers. For our model, we performed a grid search over the contrastive weight $\beta_0$ 
and warm-up duration, selecting the configuration that maximized 
validation SI-SIR performance. The final hyper-parameters are:
$\beta_0 = 3000$, 
warm-up period of 100 epochs, temperature $\tau = 0.1$ and the batch size $B = 8$.  To ensure comparability with prior work, we adopt the same data preprocessing pipeline as AV-UDiffSE+.
\begin{table}[!ht]
\centering
\caption{Ablation study of the linear projection on LRS3-NTCD.}
\label{tab:lrs3_ntcd_results_ablation}
\footnotesize
\setlength{\tabcolsep}{2.5pt}

\begin{tabular}{l c c c c c}
\toprule
Model & SI-SDR$\uparrow$ & SI-SIR$\uparrow$ & SI-SAR$\uparrow$ & PESQ$\uparrow$ & STOI$\uparrow$ \\
\midrule

w/o linear projection& 7.27&17.55 &8.74 &2.57 &	0.67 \\
w/ linear projection & \textbf{8.1} &\textbf{18.60} &\textbf{9.50} &\textbf{2.60} &\textbf{0.68} \\

\bottomrule
\end{tabular}
\end{table}
\section{Results} 
\subsection{Matched condition: TCD-DEMAND}
Under matched conditions (Table~\ref{tab:new_se_results_clean}), the proposed model improves all metrics compared to AV-DiffUSEEN. In particular, we observe around +5 dB gain in SI-SIR, indicating improved suppression of noise interference. This translates into a +2.4 dB improvement in SI-SDR, reflecting a better overall quality of reconstruction. Perceptual metrics stay close to the baseline. These results indicate that the proposed contrastive objective enhances noise interference suppression and reconstruction while maintaining perceptual quality and intelligibility. 

At low input SNR, improvements are more pronounced. The proposed method achieves a +6 dB gain in SI-SIR, +5dB in SI-SAR, a +3 dB improvement in SI-SDR and +0.06 in PESQ. This suggests increased robustness in noisy conditions, indicating that the model benefits from visual information when the acoustic modality is severely degraded.

At higher input SNR, the improvement in SI-SIR remains significant (+3.8 dB), while the remaining metrics are comparable to the baseline, still +3dB in SI-SDR, even +0.06 in PESQ. This suggests that the main contribution of the proposed method lies in interference suppression and speech quality rather than perceptual refinement when the input is already relatively clean.
\subsection{Mismatched condition: LRS3-NTCD}

Under mismatched conditions (unseen speakers, acoustic and visual environments), performance trends differ. While the proposed model still improves SI-SIR (+3.6 dB), +0.7dB in SI-SDR and +0.02 in PESQ with maintaining similar results with the remaining metrics. At low SNR, SI-SIR improvements of approximately +3.8 dB are maintained with +1dB in SI-SDR. At high SNR gains persist to about +3.5 dB with SI-SIR. Those persistent performance gains even in mismatched conditions show the model's capacity to generalize to novel and more complex data distributions. In line with~\cite{ayilo2024diffavse}, FlowAVSE exhibits a substantial performance drop.

\subsection{Visual masking at inference}
We analyze the effect of masking the visual input at inference time to assess how much our model relies on visual cues. Across all datasets, removing visual information leads to a significantly larger performance degradation for the contrastive model compared to the baseline. On TCD-DEMAND, SI-SDR decreases by about $21$~dB for the baseline, whereas our model shows a larger drop of roughly $43$~dB. For PESQ, the absolute performance drop is $0.85$ for the baseline vs. $1.56$ for our model, and for STOI it is $0.28$ vs. $0.61$, respectively. This indicates that the contrastive training objective encourages stronger cross-modal coupling, allowing the model to exploit visual information more effectively during denoising.

\subsection{Effect of the contrastive weighting factor}
\begin{figure}[t]
  \centering
  \includegraphics[width=0.9\linewidth]{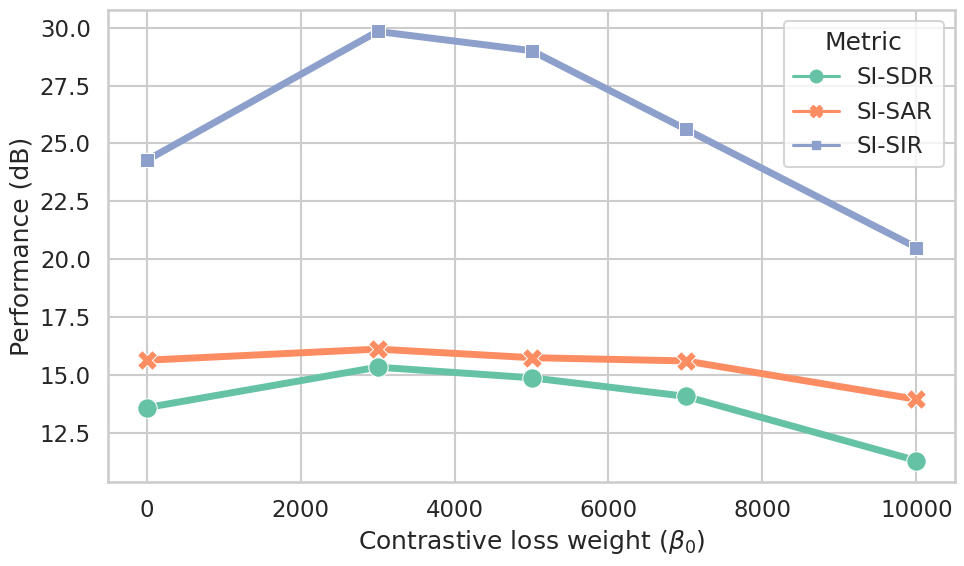}
  \caption{Effect of the contrastive loss weight $\beta_0$.}
  \label{fig:Effect of the weight}
\end{figure}

We evaluated a wide range of $\beta_0$ values, from 0 to 10{,}000. A warm-up period of 100 epochs was applied before introducing the contrastive loss, after which the trained models were evaluated on the SE task. As shown in Fig.~\ref{fig:Effect of the weight}, varying $\beta_0$ produces a consistent trend across interference suppression, artifact reduction, and overall reconstruction quality. When $\beta_0$ is too large, the contrastive objective dominates the optimization process and degrades reconstruction performance. Conversely, when $\beta_0$ is too small, the contrastive loss has limited influence and provides little benefit. Empirically, we observe the best trade-off at $\beta_0 = 3{,}000$.

\subsection{Impact of the linear projection layer}
We investigate the impact of linear projection layer before the cross-attention fusion (Fig.~\ref{fig:AV-UDiffSE with audio-visual contrastive alignment}). We observed that our model performs similarly with or without it in matched conditions. But in mismatched conditions, the linear layer makes the difference on every single metric. For example, table \ref{tab:lrs3_ntcd_results_ablation} shows a +1 dB increase in SI-SDR, SI-SIR and SI-SAR and +0.03 in PESQ; which enables our model to outperform the baseline.

\section{Conclusion} 

We studied the role of explicit cross-modal alignment in diffusion-based unsupervised audio-visual speech enhancement. During the pretraining of the visual-conditioned speech diffusion model, we augment the denoising score matching objective with a contrastive audio-visual alignment loss. Experiments show that better aligned embeddings lead to consistent gains in interference suppression, signal reconstruction, and perceptual speech quality, especially at low SNRs and under dataset mismatch.

\section{Acknowledgment}
Experiments in this work were conducted using the Grid'5000 testbed, supported by a scientific interest group hosted by Inria and including CNRS, RENATER, several universities, and other organizations (see \url{https://www.grid5000.fr}).
\section{Generative AI Use Disclosure}

All scientific content, methodology, experiments, analyses, interpretations, and conclusions were developed, verified, and approved by the authors, who take full responsibility for the contents of the paper. Generative AI tools were only used to edit and polish some portions of the manuscript.
\bibliographystyle{IEEEtran}
\FloatBarrier
\bibliography{mybib}

\end{document}